\documentstyle[preprint,aps]{revtex}
\title{Two dimensional Ising spin glasses with non-zero ordering
temperatures}  
\author{N. Lemke\footnote{Email:  {\tt lemke@lps.u-psud.fr}}
\footnote{{\it Permanent Address Instituto de F\'{\i}sica \\ 
Universidade Federal do Rio Grande do Sul  \\ C.P. 15051 -- 
91501-970 -- Porto Alegre -- RS -- Brazil \\}} 
and I. A. Campbell \footnote{Email: {\tt campbell@lps.u-psud.fr} }}
\address{Laboratoire de Physique des Solides \\
        Universit\'e Paris-Sud \\
        B\^atiment 510, Centre Universitaire,91405 Orsay C\'edex(France)  }
\date{\today}

\begin{document}

\maketitle

\begin{abstract}
      We demonstrate numerically that for Ising spins on square lattices
with ferromagnetic second neighbour interactions and random near neighbour
interactions, two dimensional Ising spin glass order with a non-zero
freezing temperature can occur. We compare some of the physical properties of
these spin glasses with those of standard spin glasses in higher
dimensions.

\end{abstract}

\vskip 1\baselineskip
\vskip 1\baselineskip
\noindent
Submitted to  {\bf Phys. Rev. Lett.}


\newpage

In dimension two, an Ising spin glass (ISG) with random near
neighbour interactions does not order above zero temperature \cite{bin1,bha1}.
Implicitly this has been taken to mean that two dimensional ISGs with
ordering at finite temperatures do not exist. Here we show that this
assumption is unfounded by exhibiting a family of 2d Ising systems with
a different set of interactions, which show spin glass ordering at finite
temperatures. We find that the physical properties of these 2d ISG systems
ressemble closely those of standard ISGs in higher dimensions.

        Consider a square lattice of Ising spins with second nearest
neighbour ferromagnetic interactions of strength $J$. This clearly consists
of two inter-penetrating but non-interacting square sublattices which will
each order ferromagnetically at the Onsager Curie temperature $T_c =2.27J$.
Now introduce random near neighbour interactions of strength $\pm\lambda J$.
At low temperatures when the spins are frozen, each sublattice will exert
effective random fields on the spins of the other sublattice. It is well
established \cite{imr1} that a 2d Ising ferromagnet breaks up into finite size
domains under the influence of a random field, however small. At low
temperatures each sublattice of the present system will then be ordered in
randomly arranged ferromagnetic domains of finite size. However in contrast
to the true 2d random field Ising system, here the ``effective random
fields'' will average to zero at high temperatures, so we can expect a
critical temperature $T_g$ separating a high temperature  paramagnetic regime
(with the autocorrelation function $q(t)$ relaxing to zero at long time $t$)
from a low temperature regime where $q(t)$ does not go to zero, corresponding
to ordering with a non-zero Edwards-Anderson order parameter. Below we will
show numerical evidence that for small and moderate $\lambda$, $T_g$ is
non-zero. In the low temperature state there is no long range ferromagnetic
order so the system is certainly not a ferromagnet; we will choose to call
the frozen state a ``spin glass'', although random systems with short range
ferromagnetic or antiferromagnetic order are often referred to as ``cluster
glasses'' in an experimental context.

        One of the established numerical techniques for determining the
ordering temperature in ferromagnets and spin glasses is through finite
size scaling using the Binder cumulant \cite{bha2}. For sets of samples of
different sizes $L$, the fluctuations of the autocorrelation function $q(t)$ in
equilibrium are recorded. The Binder cumulant at temperature $T$ is defined
as:
\begin{equation}
 g_L(T)  =\frac{1}{2} \left( 3 - \frac{\langle q^4\rangle}
                {\langle q^2\rangle^2} \right)
\end{equation}
and is dimensionless; for sets of samples with fixed $L$ values, the
series of curves $g_L(T)$ as  a function of temperature intersect at the ordering
temperature. We have carried out simulations using heat bath dynamics and
sequential spin by spin updates on samples up to $L=12$. 500 samples were
used at each size for each value of $\lambda$ and standard precautions were
taken to assure that thermal equilibrium was achieved \cite{bha2}. We show in
figure \ref{gl} the $g_L$ data for three values of $\lambda$ and it can be seen that
for the smallest  value there is a  clear intersection point. For $\lambda=0.7$
 the $g_L$ curves come together without any clear fanning out at lower temperatures, 
 but we consider that the meeting point represents the freezing
  temperature here too. For
$\lambda=1$ it is hard to say if the curves will intersect above $T=0$.
 (Very similar behaviour with almost
unobservable fanning out below $T_g$ is seen in 3d ISGs \cite{bha1,bha2}).
 The behaviour
of the freezing temperature as a function of $\lambda$ is shown in the inset of figure
\ref{Cv}; at small $\lambda$, $T_g$ extrapolates back to the Onsager $T_c$,
 and with increasing $\lambda$, $T_g$ drops  as might be expected, because for
high enough $\lambda$ we will recover the 2d random interaction near
neighbour ISG with its zero temperature ordering. It appears that the 
limiting value of $\lambda$ for non-zero $T_g$ is in the region of 1.0.

         In conventional second order transitions $g_L(T_c)$ has a universal
value for all systems in a given class; for instance for the 2d Ising
ferromagnet the magnetism Binder parameter $g_{Lm}(T_c)=0.91603$ \cite{bin2}
and we find the autocorrelation Binder parameter $g_L(T_c)=0.42$. We can note in figure (\ref{gl})
 that
$g_L(T_g)$ changes with $\lambda$ so for the present systems of 2d ISGs
universality is not obeyed. For the range of $\lambda$ that we have studied,
$g_L(T_g)$ appears to be tending regularly to  a value near 1 as $T_g$ is
tending to zero. At still higher $\lambda$ values $g_L(0)$ will presumably
tend to 0.83, which is the standard 2d $\pm J$ ISG zero temperature value
\cite{bha1,bha2}. We can note that a closely related non-random 2d Ising
 system,
having first and second neighbour interactions with regular frustation,
shows non-universality of critical exponents when a control parameter
representing the ratio of the two interaction strengths is varied \cite{min1}.
There is also strong evidence for non-universality of the critical
exponents in canonical ISGs at dimension 3 and above \cite{ber1}. We have 
not yet
estimated the values of the critical exponents for the present family of 2d
ISGs.

        Once the existence of finite ordering temperatures established, we
can carry out simulations to measure the standard physical properties for
these 2d ISGs in order to compare with those of other ISGs. In figure (\ref{Cv})
 we
show the specific heats for the same three values of $\lambda$. It can be
seen that the specific heats show broad maxima at temperatures higher than
the respective ordering temperatures, and smooth featureless behaviour
through the freezing transition. This is precisely what is observed for
standard ISGs in dimension 3 \cite{ogi1}, 4 and 5 \cite{ber2}
 and appears to be a universal
characteristic of the ISG transition below the upper critical dimension.
For the present systems the maximum of the specific heat corresponds to a
saturation of the ferromagnetic correlation length in each sublattice as
the sample is cooled; the ordering process can be thought of as consisting
of first the formation of finite size ferromagnetic domains, and then of a
freezing at a lower temperature where the domain movements are hindered by
the random near neighbour interactions.

        Another parameter accessible from simulations is the damage
spreading \cite{der1}. Damage spreading is measured numerically by using heat bath
dynamics and applying the same random update parameter at each update step
to two replicas of the same system; $D(T)$ is defined as the normalised
Hamming distance between the two replicas (i.e. the fraction of spins
having opposite orientations) at long times when this protocol is applied.
While ferromagnets have $D(T) =0$ down to $T_c$, standard ISGs have been found
to show non-zero $D(T)$ until temperatures well above $T_g$ \cite{der1,cam1};
 it has been
pointed out that for ISGs in dimension 3 and 4, as the temperature is
reduced $D(T)$ tends to 1/2 at a temperature very close to $T_g$ \cite{cam1}.
 We have
studied damage spreading in the present 2d ISGs. Two random initial
replicas of each 2d system were chosen; these replicas were annealed
independently to thermal equilibrium at each temperature, and then the
damage spreading procedure was applied for long times. The results, figure
(\ref{dam}), show that the damage spreading temperature $T_D $ where $D(T)$ 
first becomes
non-zero is much higher than $T_g$, and that in each case on cooling $D(T)$
tends to 1/2. For $\lambda=0.5$ this temperature is indistinguishable from the
 Tg we have
estimated from the Binder cumulant data. $D(T)$ then remains equal to 1/2 for
lower temperatures. The  behaviour is similar but less clear cut for the 
other two values of $\lambda$ (the lowest temperature points may be slightly low
because of the very long times needed to achieve complete 
thermal equilibrium). The general behaviour of the damage spreading parameter
appears to be the same in these 2d ISGs as  in ISGs in dimensions 3 and 4.
The use of the empirical criterion $D(T_g) =1/2$ gives a convenient method
to check $T_g$ values.

        An advantage of the present systems over conventional ISGs is that
``snapshots'' of the instantaneous configurations of the spins are
informative because we are in dimension 2.
 We show in figure (\ref{snap}) two snapshots of the same
$\lambda=0.5$
system at temperature $T=1.5$, well below $T_g$. The images were produced by
slowly cooling two independent random replicas of the same  system by
successive steps to the final temperature. As could be expected there are
four types of domains : if we call one sublattice A and the other B then
the domains are A up B up (white), A up B down (white/black checkerboard),
A down B up (black/white) and A down B down (black). Here the domain sizes
are typically 25 by 25 spins. (The simulations confirm that the typical
domain size below $Tg$ gets smaller as $\lambda$ is increased). For this sample
 at
this temperature the two configurations are almost frozen; averaging over a
further 10000 Monte Carlo Steps per spin (MCS) only blurs the domain
frontiers somewhat. To a good approximation the images then represent two
Gibbs states of the system. By inspection it can be seen that the two
states are neither quasi-identical, nor are they quasi-mirror images of
each other with all the spins reversed. The detailed structures of the two
states are quite different suggesting that there are an infinity (or at
least a large number) of alternative Gibbs states for this ISG. This
conclusion appears to be incompatible with the two ground state image which
has been proposed heuristically for ISGs \cite{fis1} but the phase space could
ressemble that of the Parisi solution of the infinite range SK model 
\cite{par1}.

        In conclusion, we have demonstrated that 2d ISG systems with
non-zero freezing temperatures exist. We have shown data for a certain
number of parameters for a new family of 2d ISGs where we have found that
the observed behaviour is very similar to that seen in simulations on
standard ISGs in dimension 3. A more comprehensive study of the 2d ISGs and
a detailed comparison with results from higher dimensions should establish
what properties are always necessarily associated with spin glass ordering
and should help understanding of the spin glass phenomenon in general. As
far as experimental comparisons are concerned, it might be worth
reconsidering the data on 2d ISG materials without making the a priori
assumption that the ordering temperature is necessarily zero.

        Finally, the general philosophy behind the present study may help
to resolve a long standing paradox. Real  three dimensional spin
glass materials  where the spins can be considered to be basically 
Heisenberg show
well defined non-zero freezing temperatures; on the other hand in numerical
work and on theoretical grounds it appears that Heisenberg spins with
random near neighbour interactions should not show a freezing transition at
a finite temperature \cite{oli1} (but it should be noted that with vector
 spins
chirality should also play a role \cite{kaw1} and that a simulation of a Heisenberg spin 
glass with RKKY interactions shows a finite $T_g$ \cite{mat1}). 
Now we have just seen that if
 we
have a set of random near neighbour interactions only, a 2d Ising system
has no finite temperature ordering,  but the system can be  a spin glass
with a finite $T_g$ if there are both random and non-random interactions.
Perhaps the same rule applies to 3d Heisenberg systems. It may well be that
all real life Heisenberg spin glasses have both random and non-random
interactions (and so in a sense might be classified as ``cluster glasses'' 
\cite{nie1}).
Thus even in the canonical CuMn alloys, the short range magnetic ordering
which has been observed with neutrons \cite{cab1} may be essential for the
 spin
glass transition to occur at a finite temperature.

\section*{Acknowledgements}
We would like to thank Lorenzo Bernardi for his contribution for programming.
The numerical calculations were carried out thanks to a time allocation
provided by the Institut du Developpement et des Ressources en Informatique
Scientifique (IDRIS). N.L. is supported by a CNPq scholarship.

\newpage

\begin{figure}
\caption{The measured Binder Cumulants as a function of size L and temperature
 T for $\lambda=0.5, 0.7$ and 1.0 respectively.
}
\label{gl}
\end{figure}

\begin{figure}
\caption{ The specific heat averaged over samples with $L=40$ for $\lambda=0.5,
0.7$ and 1.0 (circles, squares and triangles respectively). \\ Inset $T_g$ as
 a function of $\lambda$} 
\label{Cv}
\end{figure}

\begin{figure}
\caption{ The damage spreading parameter D(T) as a function of temperature
(see text) for $\lambda$ =0.5 (triangles), 0.7 (circles) and 1.0 (squares) for
$40\times 40$ spins.}
\label{dam}
\end{figure}

\begin{figure}
\caption{Snapshots of two replicas A,B of the same $100\times 100$ spin system 
with $\lambda=0.5$ at T=1.5 after 10000 mcs anneal. The two replicas were 
cooled independently from two initially randomly chosen configurations On 
each site black is up white is down. The third frame is the difference
between the two configurations A and B; black: $S_i^A=S_i^B$, white $S_i^A=-
S_i^B$.}

\label{snap}
\end{figure}

\newpage
\begin{thebibliography}{40}

\bibitem{bin1}{ K. Binder and A.P. Young}, {\it Rev. Mod. Phys.} {\bf 58} , 801 (1986)

\bibitem{bha1}{ R.N. Bhatt and A.P. Young}, {\it Heidelberg Colloquium on 
Glassy Dynamics},
edited by J.L. van Hemmen and I. Morgenstern (Springer Verlag Berlin) 1987

\bibitem{imr1}{ Y. Imry and S.K. Ma}, {\it Phys. Rev. Lett.} {\bf 35}, 1399 (1975)

\bibitem{bha2}{ R.N. Bhatt and A.P. Young}, {\it Phys. Rev. B} {\bf 37}, 5606 (1988)

\bibitem{bin2}{ K. Binder},{\it Z. Phys. B} {\bf 43}, 119 (1981)

 G. Kamieniarz and H.W.J. Bl\"ote, {\it  J. Phys. A} {\bf 26}, 201 (1993)
 
\bibitem{min1}{ K. Minami and M. Suzuki}, {\it Physica A}{\bf 195}, 457 (1993)

\bibitem{ber1}{ L. Bernardi and I.A. Campbell}, {\it Phys. Rev. B}
 {\bf 52}, 12501 (1995)

\bibitem{ogi1}{ A.T. Ogielski}, {\it Phys. Rev. B} {\bf 32}, 7384 (1985)

\bibitem{ber2}{ L. Bernardi and I.A. Campbell}, unpublished

\bibitem{der1}{ B. Derrida and G. Weisbuch}, {\it Europhys. Lett.}
{\bf 4}, 657 (1987) 

 B. Derrida, {\it Phys. Rep.} {\bf 184}, 207 (1989)
 
\bibitem{cam1}{ I.A. Campbell}, {\it Europhys. Lett.} {\bf 21}, 959 (1993)

\bibitem{fis1}{ D.S. Fisher and D.A. Huse}, {\it Phys. Rev. Lett.} 
{\bf 56}, 1601 (1986)

\bibitem{par1}{ G. Parisi}, {\it Phys. Rev. Lett.} {\bf 50}, 1946 (1983)


 { E. Marinari, G. Parisi, J. Ruiz-Lorenzo and F. Ritort}, 
 {\it Phys. Rev. Lett.} {\bf 76}, 843 (1996)
 
\bibitem{oli1}{ J.A. Olive, A.P. Young and D. Sherrington}, 
{\it Phys. Rev. B} {\bf 34}, 6341 (1986)

 { H. Yoshino and H. Takayama}, {\it Europhys. Lett.} {\bf 22}, 631 (1993)
 
\bibitem{kaw1}{ H. Kawamura}, {\it J. Phys. Soc. Japan} {\bf 64}, 26 (1995)

\bibitem{mat1} { F. Matsubara and M. Iguchi}, {\it Phys. Rev. Lett.}, {\bf 68},
3781, 1992.

\bibitem{nie1}{Ordering in Cluster Glasses is discussed by Th. M. Nieuwenhuizen and C. N. A. van Duin}, unpublished.  


\bibitem{cab1}{ J.W. Cable, S.A. Werner, G.P. Felcher and N. Wakabayashi}, 
{\it Phys. Rev. B} {\bf 29}, 1268 (1984)


  
\end{thebibliography}
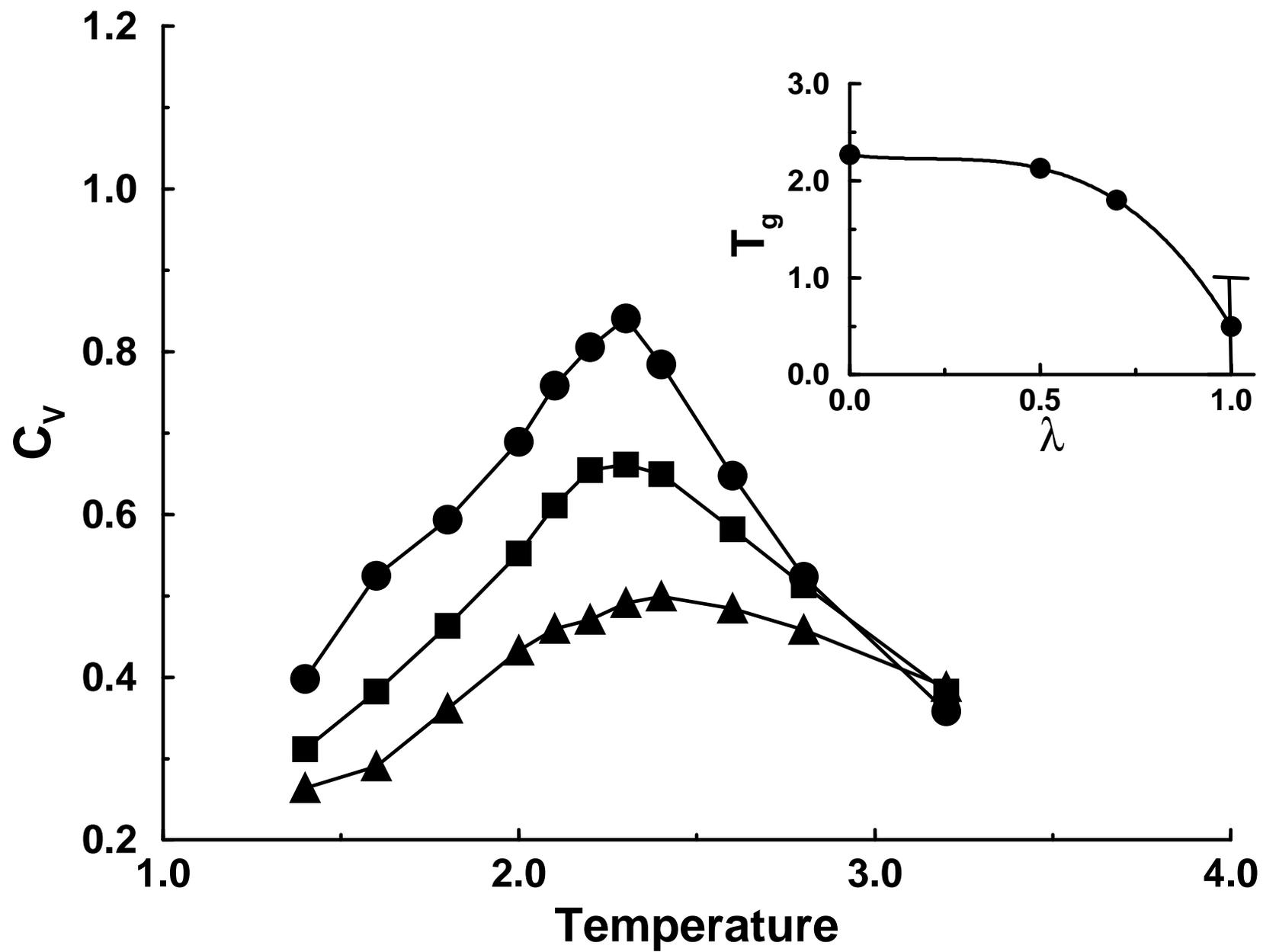
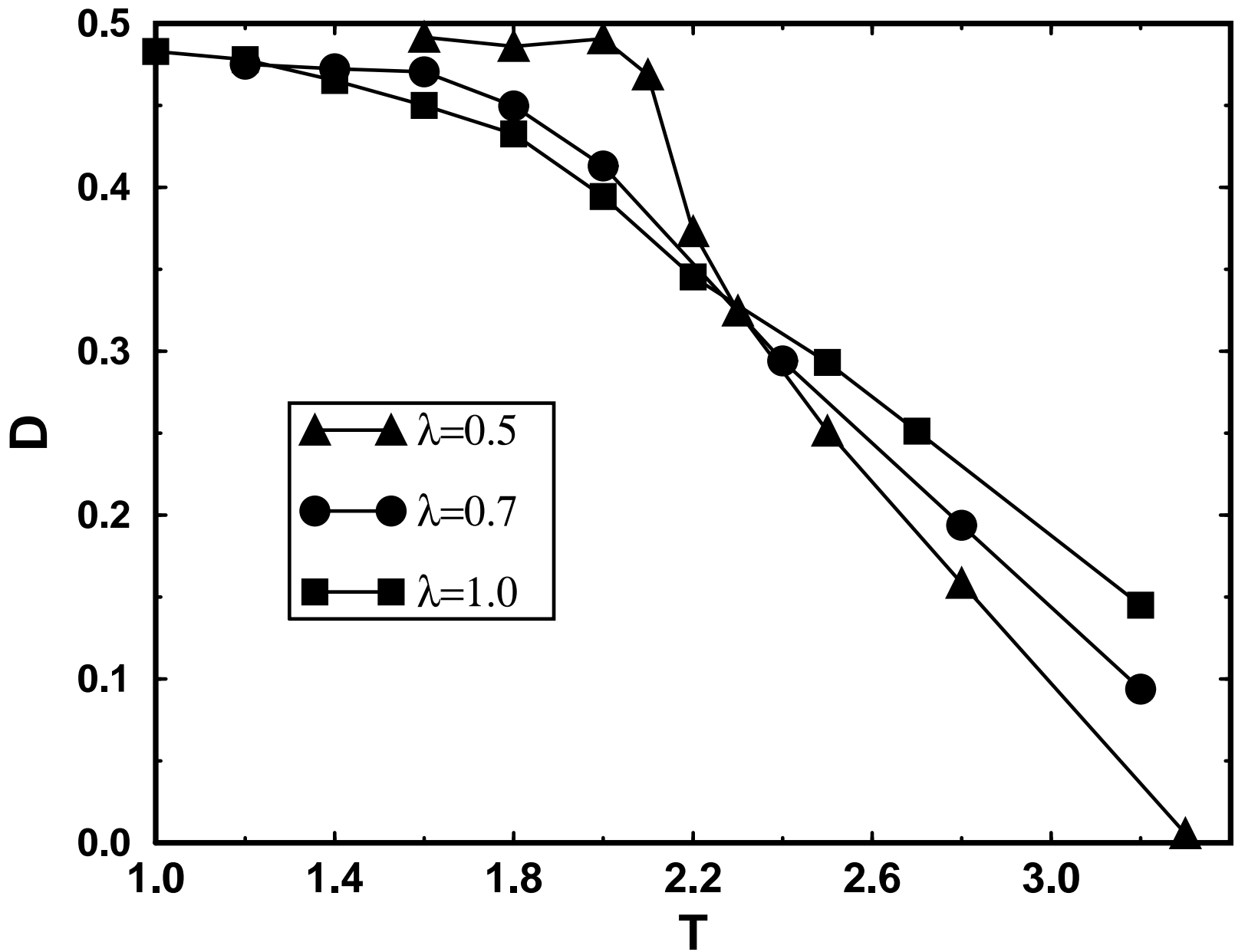
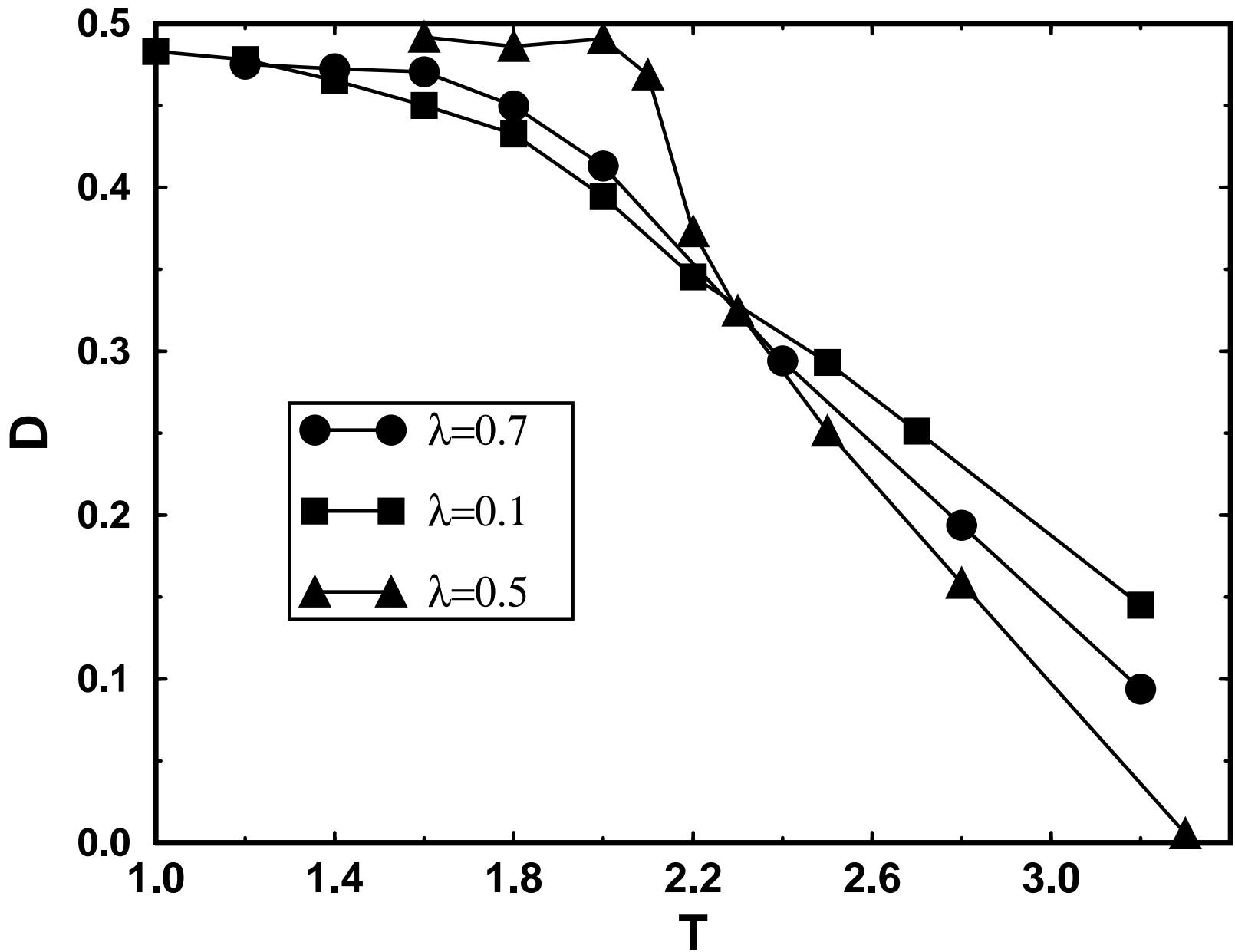
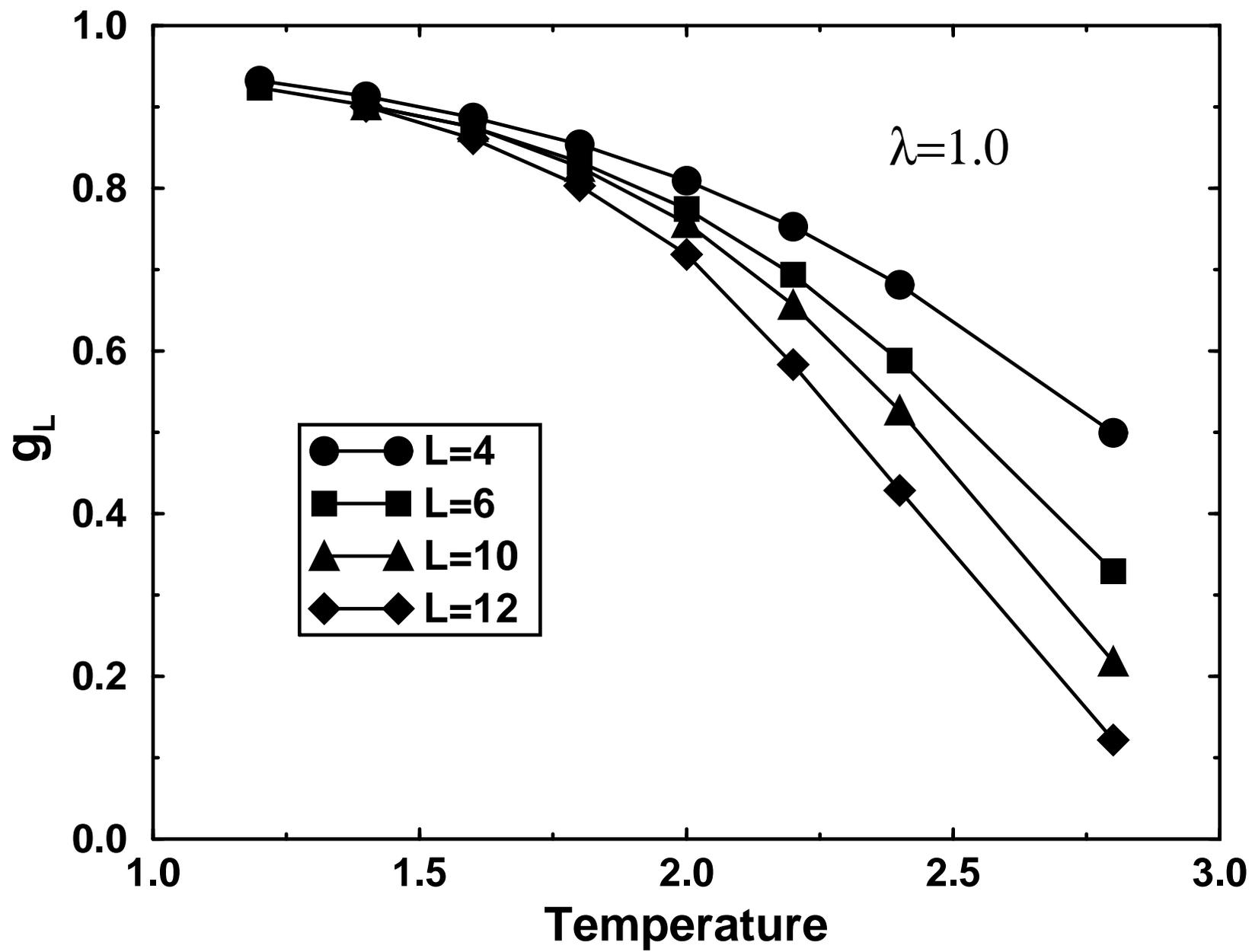
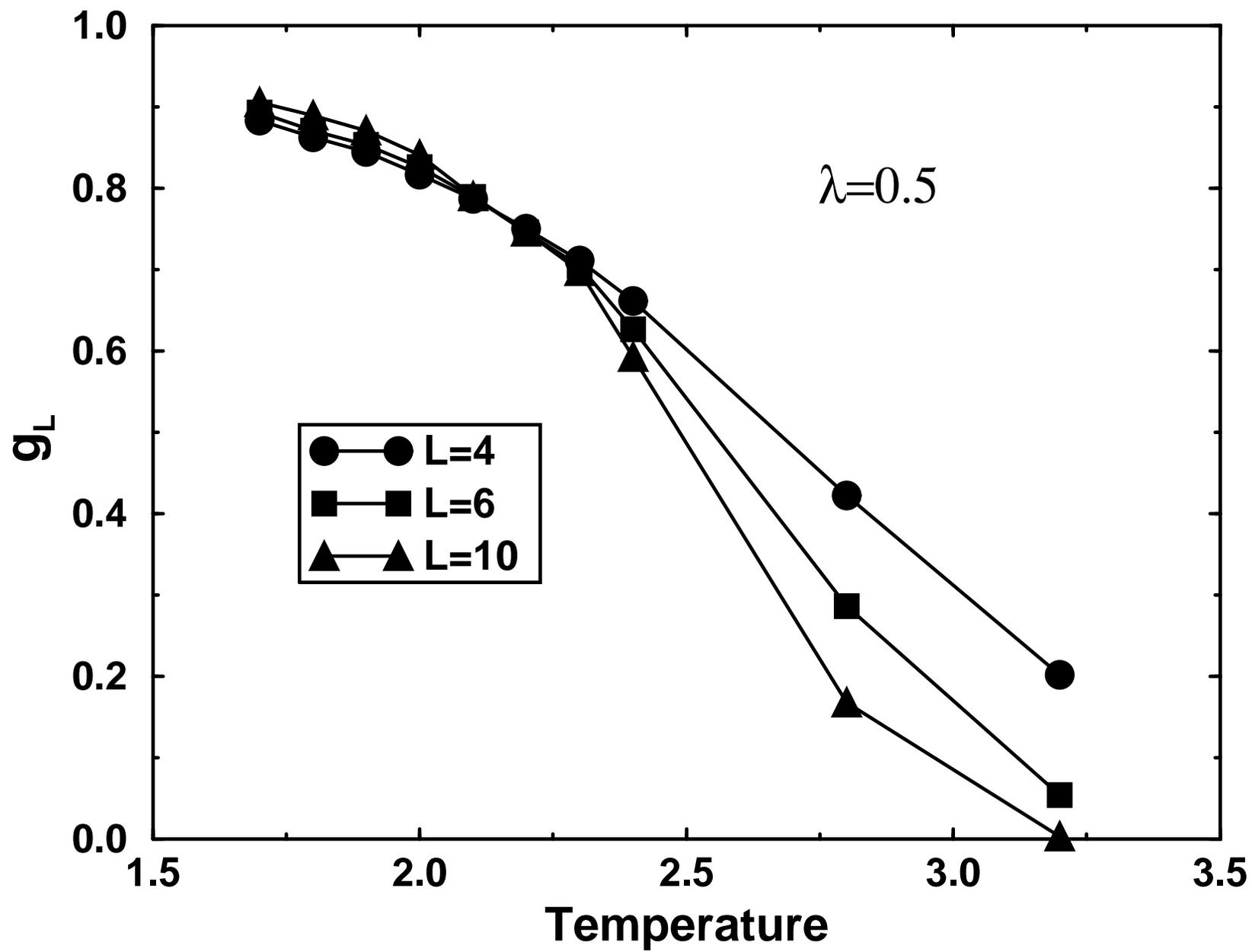
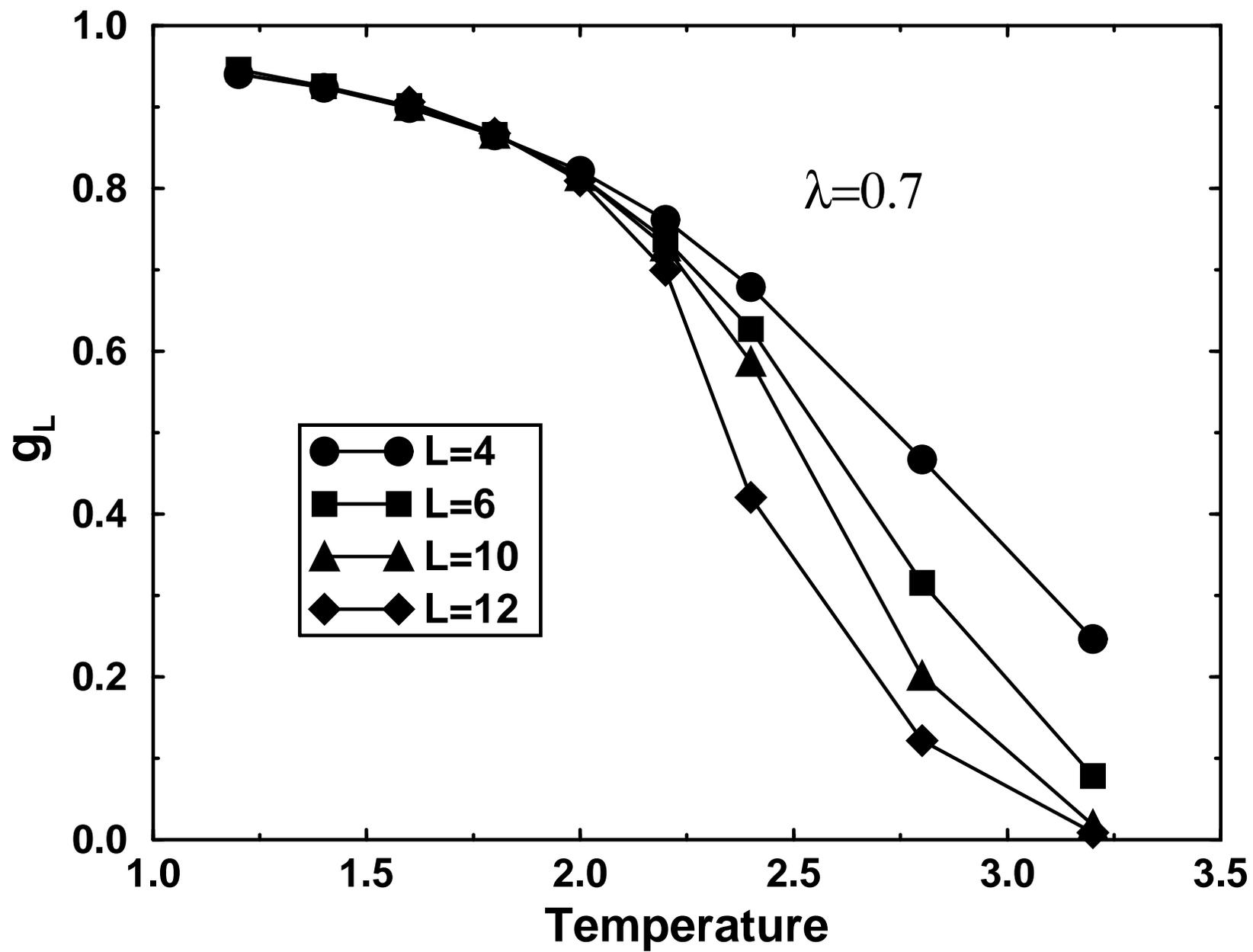
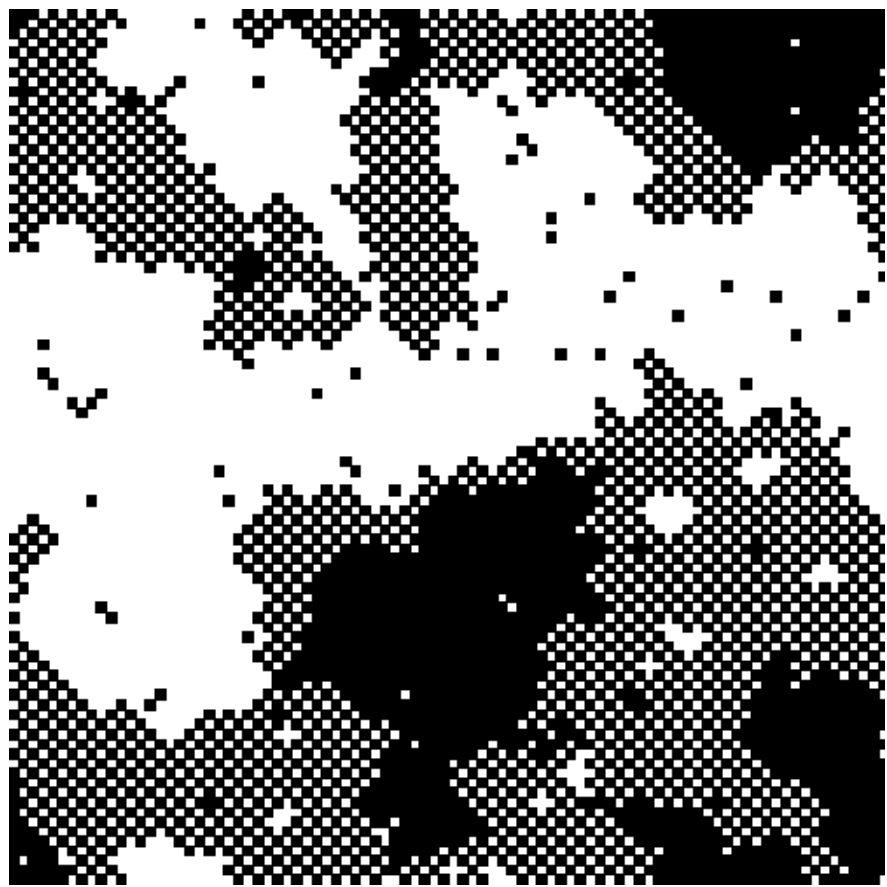
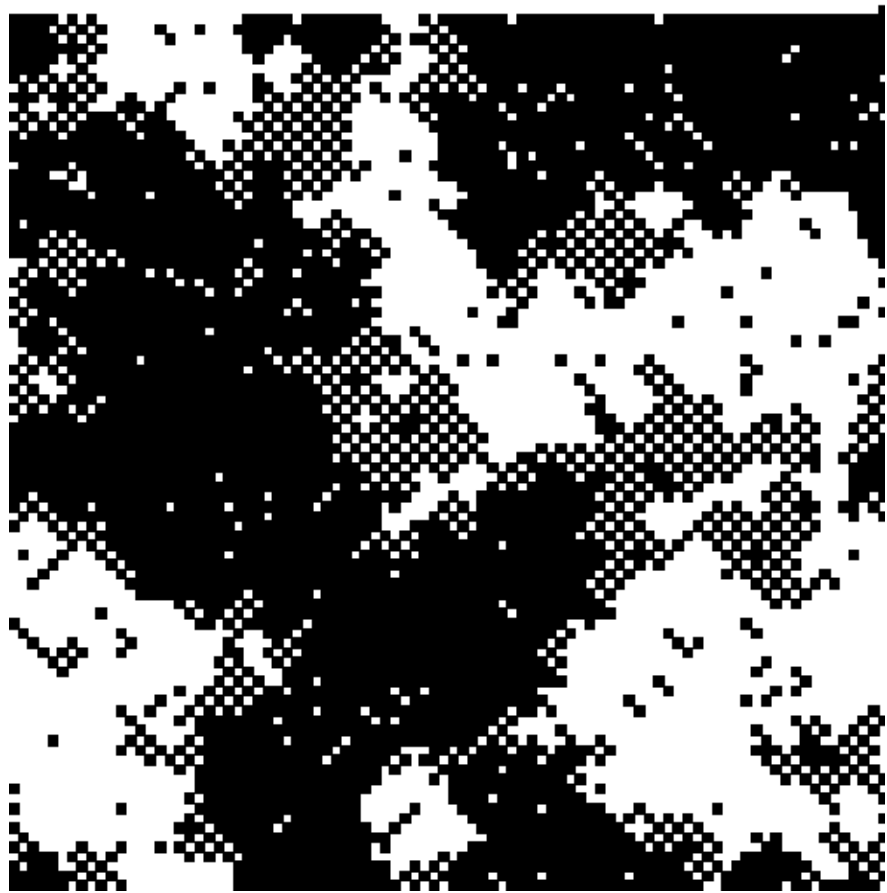
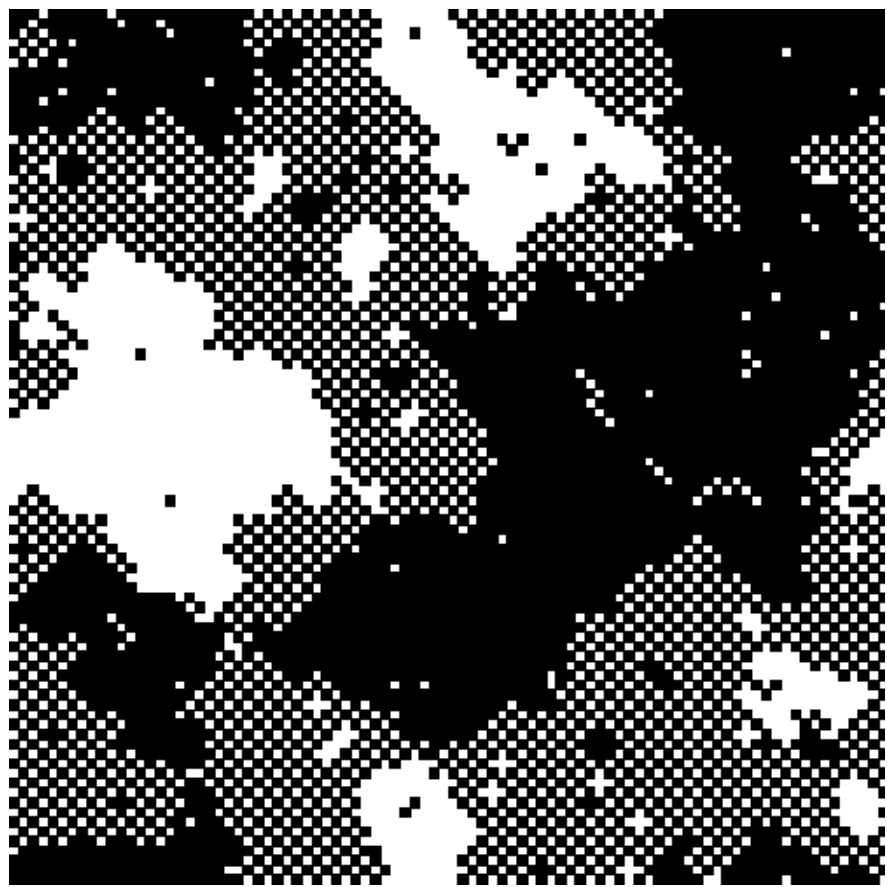
\end{document}